\documentclass[10pt]{iopart}
\usepackage[english]{babel}
\usepackage{hyperref}
\usepackage{amssymb}
\usepackage{graphicx}
\usepackage{comment}
\usepackage[titletoc,toc,title]{appendix}
\usepackage[colorinlistoftodos]{todonotes}
\usepackage[utf8]{inputenc}
\usepackage{siunitx}
\usepackage{bm}
\usepackage{cite}
\usepackage{epstopdf}
\usepackage{float}

\expandafter\let\csname equation*\endcsname\relax
\expandafter\let\csname endequation*\endcsname\relax 
\usepackage{amsmath}
\usepackage{leftidx}
\usepackage{doi}

\usepackage[toc,acronym]{glossaries}
\newacronym{GPE}{GPE}{Gross-Pitaevskii equation}
\newacronym{BKT}{BKT}{Berezinskii-Kosterlitz-Thouless}
\newacronym{1D}{1D}{one-dimensional}
\newacronym{2D}{2D}{two-dimensional}
\newacronym{3D}{3D}{three-dimensional}
\newacronym{RF}{RF}{radiofrequency}
\newacronym{BEC}{BEC}{Bose-Einstein condensate}
\newacronym{PSD}{PSD}{phase-space density}
\newacronym{TOF}{TOF}{time-of-flight}
\newacronym{ROI}{ROI}{region-of-interest} 
\newacronym{MWI}{MWI}{Matter-Wave Interference}
\newacronym{DMD}{DMD}{Digital Micromirror Device}

\DeclareSIUnit\gauss{G}

\newcommand*{\rf}[0]{RF}
\newcommand*{\Rb}[1]{\ensuremath{\mathrm{^{#1}Rb}}}
\newcommand{\Fo}{$F=1$}
\newcommand{\Ft}{$F=2$}

\begin{document}

\title{Coherent splitting of two-dimensional Bose gases in magnetic potentials}

\author{A.~J.~Barker, S.~Sunami, D.~Garrick, A.~Beregi, K.~Luksch, E.~Bentine and C.~J.~Foot}

\address{Clarendon Laboratory, University of Oxford,
Oxford OX1 3PU, United Kingdom}
\ead{christopher.foot@physics.ox.ac.uk}

\begin{indented}
\item[]\today
\end{indented}

\begin{abstract}
	
	Investigating out-of-equilibrium dynamics with two-dimensional (2D) systems is of widespread theoretical interest, as these systems are strongly influenced by fluctuations and there exists a superfluid phase transition at a finite temperature.
	In this work, we realise matter-wave interference for degenerate Bose gases, including the first demonstration of coherent splitting of 2D Bose gases using magnetic trapping potentials.
	We improve the fringe contrast by imaging only a thin slice of the expanded atom clouds, which will be necessary for subsequent studies on the relaxation of the gas following a quantum quench.
	
\end{abstract}

\maketitle

\section{Introduction}

A pioneering early experiment on Bose-Einstein condensates (BECs) was the demonstration of \gls{MWI} when two clouds of ultracold atoms overlap after being released from their confining potentials~\cite{Andrews1997}.
This process is an atomic analogue of the optical interference observed in Young's double-slit apparatus~\cite{YoungsSlits}, and is a powerful technique to extract information about the relative phase of the wave functions~\cite{Kasevich2002,Bongs2004,Navon2015,PethickSmith,Hadzibabic2006}. 
Moreover, the dynamics of ultracold atoms occur on experimentally observable timescales, thus these systems are ideal environments to study out-of-equilibrium phenomena in real time~\cite{Langen2015a, Prufer2018,Erne2018,Glidden2020}.

The access to the phase distributions provided by \gls{MWI} permits the investigation of phase fluctuations which are intrinsic to low-dimensional quantum systems~\cite{Hofferberth2007a,Hofferberth2008,Langen2015a,Langen2013} and prevent the formation of true long-range order~\cite{Mermin1966}.
For \gls{2D} systems, there exists a \gls{BKT} transition to a superfluid phase possessing quasi-long-range order~\cite{Kosterlitz1973,Berezinskii1971a}, which has been observed in ultracold gases~\cite{Hadzibabic2009,Hadzibabic2006,Chomaz2015,Choi2013,Merloti2013,Luick2019} amongst other contexts~\cite{Amo2009,Snoke2002,Butov2007}.
\gls{2D} systems are therefore a rich environment in which fluctuations and a superfluid phase transition at finite-temperature can be observed.
Furthermore, there is considerable theoretical interest in probing the out-of-equilibrium dynamics of the \gls{BKT} phase close to the critical point.
Reference~\cite{Mathey2017} proposes a scheme whereby a \gls{2D} Bose gas is split into two parallel daughter clouds.
The splitting process quenches the density, leaving the daughter clouds in a superheated state, characterised by a \gls{PSD} below the critical value required for superfluidity. 
The gases subsequently cross the \gls{BKT} phase transition dynamically, from superfluid to normal fluid, analogous to the reverse of the Kibble-Zurek mechanism which describes the onset of structure in many contexts~\cite{Kibble1976,Mathey2010,Navon2015,Beugnon2017}.

In this work, we demonstrate significant progress towards studies of out-of-equilibrium dynamics in \gls{2D} systems, including the first demonstration of coherent splitting of \gls{2D} gases confined by magnetic potentials. 
These traps are formed by dressing atoms with multiple \rf{} (radiofrequency) fields~\cite{Harte2018,Luksch2019,Barker2020}, whose amplitudes can be controlled to tune each trapping well between \gls{3D} and quasi-\gls{2D} confinement~\cite{Merloti2013,Perrin2017}; their extremely low heating rates also make them ideally suited to studies of time-dependent processes. 

This paper is structured as follows: in section~\ref{sec:ExperimentalSequence}, we present experimental methods to produce \glsplural{BEC} in magnetic double-well potentials formed using multiple \rf{} fields.
In section~\ref{sec:Analysis}, we discuss analysis methods applied to the \gls{MWI} pattern of \glsplural{BEC} released from a double-well potential. 
In section~\ref{sec:FringeResults}, we verify the expected nature of the interference fringes as we change the \gls{TOF} and the spatial separation between the trapping wells. 
In section~\ref{sec:Coherence}, we analyse the properties of the \gls{MWI} fringes and determine the repeatability of the relative phase for daughter clouds in the \gls{3D} regime.
In section~\ref{sec:2D}, we increase the confinement strength provided by the double-well potential, which reduces the effective dimensionality of the trapped gas to two dimensions.
We also demonstrate coherent splitting and other properties of the fringes necessary for studies of subsequent time-dependent dynamics.
We control our potential to balance the population of atoms in each well to maximise the contrast of the \gls{MWI} fringes. 
In section~\ref{sec:Tomo}, we discuss a method of selective imaging using a \gls{DMD} that improves the contrast of \gls{MWI} patterns.
Finally, in section~\ref{sec:Conclusion}, we discuss the future experiments that are possible with our \gls{2D} double-well potential.

\section{Experimental Methods}			
\label{sec:ExperimentalSequence}

Our experimental sequence follows that presented in \cite{Harte2018}.
A BEC of approximately $1.5 \times 10^5$ atoms of \Rb{87} in the lower hyperfine state is first loaded into a single-well \rf{}-dressed potential formed by a single-frequency \rf{} field.
The double-well potential is then produced using an \rf{} field with three frequency components; a detailed explanation of this technique is presented in reference~\cite{Harte2018}.
The separation between the potential minima of the double well depends on the magnetic quadrupole field gradient and the frequency difference between the three \rf{} components~\cite{Luksch2019,Barker2020}.
We first consider a double well formed by \rf{} components with frequencies \SI{1.8}{\mega\hertz}, \SI{2.0}{\mega\hertz} and \SI{2.2}{\mega\hertz}, with a quadrupole gradient of \SI{165}{\gauss\per\centi\metre}.
The \SI{1.8}{\mega\hertz} and \SI{2.2}{\mega\hertz} components define the upper and lower well, respectively, of the double-well potential aligned along the $z$-axis.

\begin{figure}
	\centering
	\includegraphics[width = \linewidth]{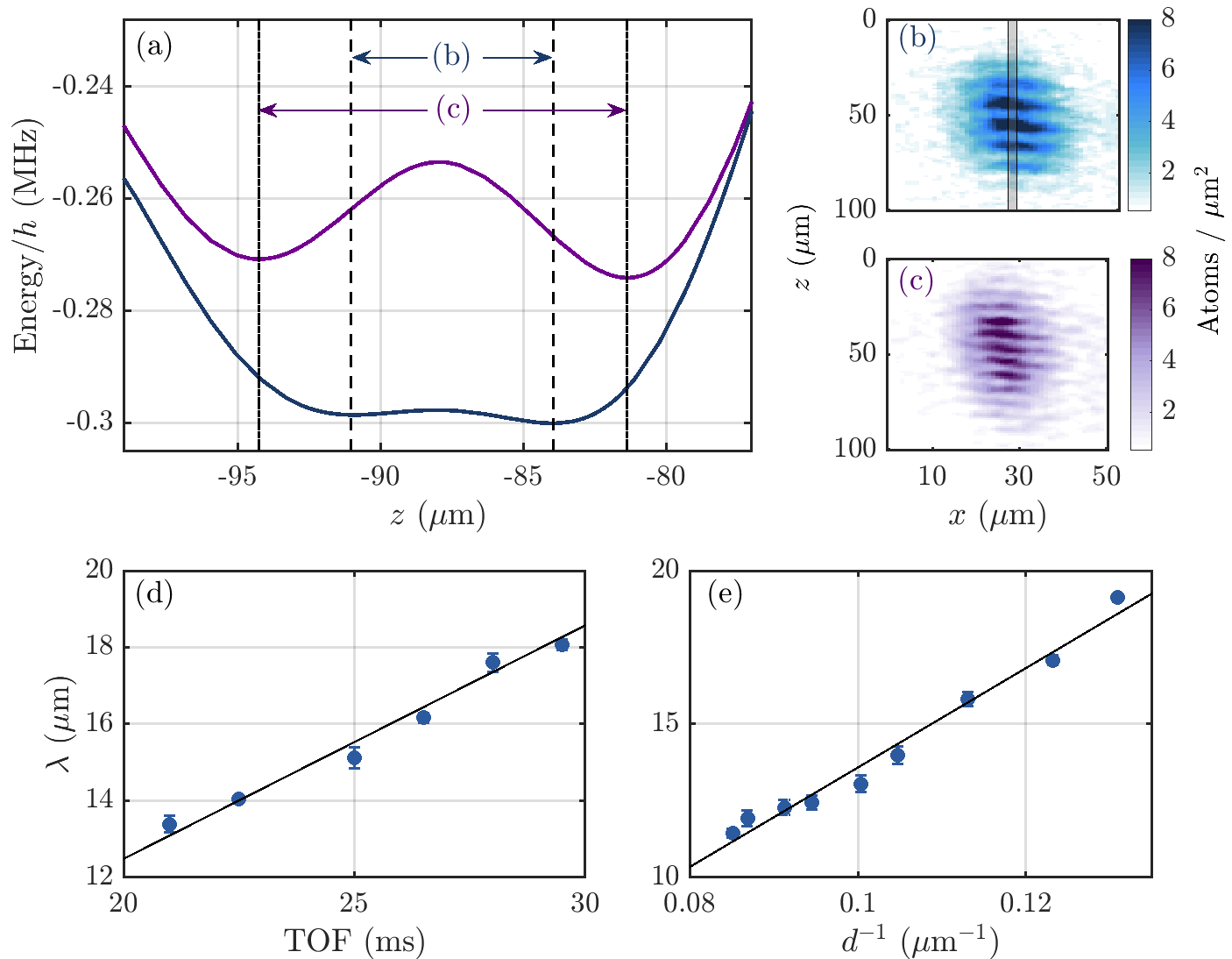}
	\caption[]{\gls{MWI} of \glsplural{BEC} in double wells. (a) Multiple-\rf{}-dressed eigenenergies for \rf{} components at \SI{1.8}{\mega\hertz}, \SI{2.0}{\mega\hertz} and \SI{2.2}{\mega\hertz}. The amplitude of the \SI{2.0}{\mega\hertz} field can be controlled to raise or lower the central barrier and to modify the spatial separation between the wells. Barrier amplitudes of \SI{314}{\milli\gauss} and \SI{185}{\milli\gauss} realise well separations of \SI{7.1}{\micro\metre} (b, blue) and \SI{12.9}{\micro\metre} (c, purple), respectively. Dashed and dashed-dotted lines mark positions of the potential minima for the configurations corresponding to (b) and (c), respectively. (b) \& (c): Absorption images of the \gls{MWI} produced by \glsplural{BEC} initially confined in the potentials (b) and (c) after \SI{25}{\milli\second} of \gls{TOF}. The grey box in (b) indicates the strip of pixels used in further analysis of the fringe spacing and phase, as discussed in section~\ref{sec:Analysis}. (d) \& (e): Variation of the spacing $\lambda$ of \gls{MWI} fringes as a function of \gls{TOF} (d) and inverse well separation (e), which show the linear relationships expected from equation~\ref{eq:wavel}. We determine the well separation from a Floquet theory-based numerical simulation of the trapping potential.}
	\label{fig:Schematic}
\end{figure}

Finer control of the separation between the two wells is implemented via the amplitude of the \SI{2.0}{\mega\hertz} component, which defines the `barrier' of the double-well potential.
This is demonstrated in figure~\ref{fig:Schematic}~(a), which depicts the potential formed with a barrier \rf{} amplitude of \SI{314}{\milli\gauss} (b, blue) or \SI{185}{\milli\gauss} (c, purple), to realise spatial separations between the minima of the potential wells of \SI{7.1}{\micro\metre} (dashed) and \SI{12.9}{\micro\metre} (dashed-dotted), respectively.
The potential can also be tilted by adjusting the amplitudes of the \rf{} fields which form the upper and lower wells in order to control the population of atoms in each well~\cite{Harte2018}.

\section{Interference Fringes}
\label{sec:Analysis}

We release the clouds by turning off the \rf{} fields while leaving the quadrupole magnetic field on, which projects the atoms in each well into the Zeeman substates, labelled by quantum numbers $m_F$~\cite{Bentine}.
The $m_F=0$ components of each trapped gas are unaffected by the magnetic field and fall freely. 
As shown in figure~\ref{fig:Schematic}~(b)~\&~(c), an interference pattern in the density distribution is observed as the $m_F=0$ components from each well overlap.
Similar to Young's double-slit experiment, the dependence on spatial separation between the wells is clear from these images, which correspond to initial separations provided by potentials (b) and (c), respectively.

The expected functional form of the interference pattern can be understood by arguments presented in references~\cite{PethickSmith,Andrews1997}.
In the \gls{3D} regime, Bose-condensed gases are characterised by a global phase and a wave function of the form $\psi(\mathbf{r},t)\sim\sqrt{n(\mathbf{r},t)}~\mathrm{exp}(i\phi(t))$, where $\mathbf{r}$ and $t$ are position and time, respectively.
In our double-well system, the combined wave function can be expressed $\Psi(\mathbf{r},t) = \psi_u(\mathbf{r},t) + \psi_l(\mathbf{r},t)$, where $\psi_{u,l}(\mathbf{r},t)$ refer to the wave functions of the \glsplural{BEC} in the upper and lower wells, respectively.
Once the atomic clouds are released from the potential, they overlap and interfere.

For \gls{TOF} durations $t_{\mathrm{TOF}}\gg1/f_z$, where $f_z$ is the axial trap frequency, the size of the clouds along the $z$-axis is much greater than their initial extent, and the two clouds have a high degree of overlap.
For a given position along the $x$-axis, the density distribution along a line parallel to the $z$-axis is thus described by
\begin{equation}
\label{eq:fit}
n(z,t_{\mathrm{TOF}};x) = g(z,t_{\mathrm{TOF}};x) \left( 1+ A~\mathrm{cos} \left( \frac{2 \pi z}{\lambda} + \Phi_x \right) \right) \hspace{5pt} , 
\end{equation}
where $g(z,t_{\mathrm{TOF}};x)$ is an envelope function, $A$ is the fringe contrast, and $\Phi_x = \phi_u - \phi_l$ is the local relative phase between the clouds. 
$\lambda$ is the fringe spacing, which is given by
\begin{equation}
\label{eq:wavel}
\lambda = \frac{ht_{\mathrm{TOF}}}{Md} \hspace{5pt} ,  
\end{equation}
where $M$ is the atomic mass and $d$ is the spatial separation between the wells. 
Both clouds are assumed to have the same trapping conditions and initial spatial extent, as characterised by the oscillation frequencies for atoms within each potential well. 
In reality, these trap frequencies in the two wells are not equal because the amplitudes of the \rf{} fields which produce the trapping wells must differ to counteract the difference in gravitational potential energy~\cite{Bentine2017}.
Throughout experiments described in this paper, the relative difference in trap frequencies remains $<15\%$, and does not lead to any significant differences between the expansion dynamics of the two clouds.
The simple model ignores the incoherent thermal component, which appears as a background distribution that reduces the contrast of the fringes~\cite{Rath2010}.
Using these tools, we now investigate the interference of \glsplural{BEC} in the \gls{3D} regime using double-well potentials.

\section{Interference of three-dimensional BECs}			
\label{sec:FringeResults}

First, we investigate the scaling of the \gls{MWI} fringe spacing as we vary $d$ and $t_{\mathrm{TOF}}$.
Experimental measurements were performed with approximately $3 \times 10^4$ atoms in each well.
Figure~\ref{fig:Schematic}~(b) illustrates the vertical strip of pixels centred on the interference pattern from which we extract $n(z,t_{\mathrm{TOF}})$, which is further analysed to determine the fringe spacing $\lambda$ using equation~\ref{eq:fit}. 
Figure~\ref{fig:Schematic}~(d) illustrates the expected linear relationship between $t_{\mathrm{TOF}}$ and the fitted $\lambda$, as described by equation~\ref{eq:wavel}.
From the gradient of the fitted curve, we extract the well separation as 7.4(0.4)~\si{\micro\metre}, which is consistent with a numerical simulation of the eigenenergies using the Floquet method which predicts a separation of \SI{7.1}{\micro\metre}~\cite{Bentine2020}.

We now vary the spatial separation between the potential minima over the range 11.6-7.0~\si{\micro\metre} by adjusting the amplitude of the \SI{2.0}{\mega\hertz} field which forms the barrier of the double-well potential.
The well separation is determined from a Floquet theory-based numerical simulation of the multiple-\rf{}-dressed potential for the specific \rf{} amplitudes in each experimental cycle, as measured by calibrated antennae situated in the apparatus~\cite{Bentine2020}.
Figure~\ref{fig:Schematic}~(e) illustrates the linear relationship between fringe spacing and inverse well separation $d^{-1}$, as expected from equation~\ref{eq:wavel}.

Non-zero interparticle interactions, as well force gradients arising due to the non-linear Zeeman effect which are not considered in the model presented in section~\ref{sec:Analysis}, may influence the expansion dynamics of the trapped gases and thus lead to a systematic shift in the measured fringe spacing.
From numerical simulations of the expansion dynamics, and for the choices of experimental parameters used here and throughout this work, we find these effects lead to shifts of less than 10\% from the model introduced earlier (equation~\ref{eq:wavel}).

We note the interference fringes are tilted by a small angle with respect to the $z$-axis, which may occur due to dipolar oscillations between the initial gases or an asymmetry in the trapping potential.
While a strip of pixels is used to determine the properties of the interference pattern here, this would introduce a source of systematic error if the interference fringes were more severely tilted. 
An approach outlined in reference~\cite{Polkovnikov2006} is to fit the integrated interference pattern, where the optimal line of integration is determined by maximising the fringe contrast of the integrated pattern. 

\section{Coherence}
\label{sec:Coherence}

\begin{figure}
	\centering
	\includegraphics[width = 0.9\linewidth]{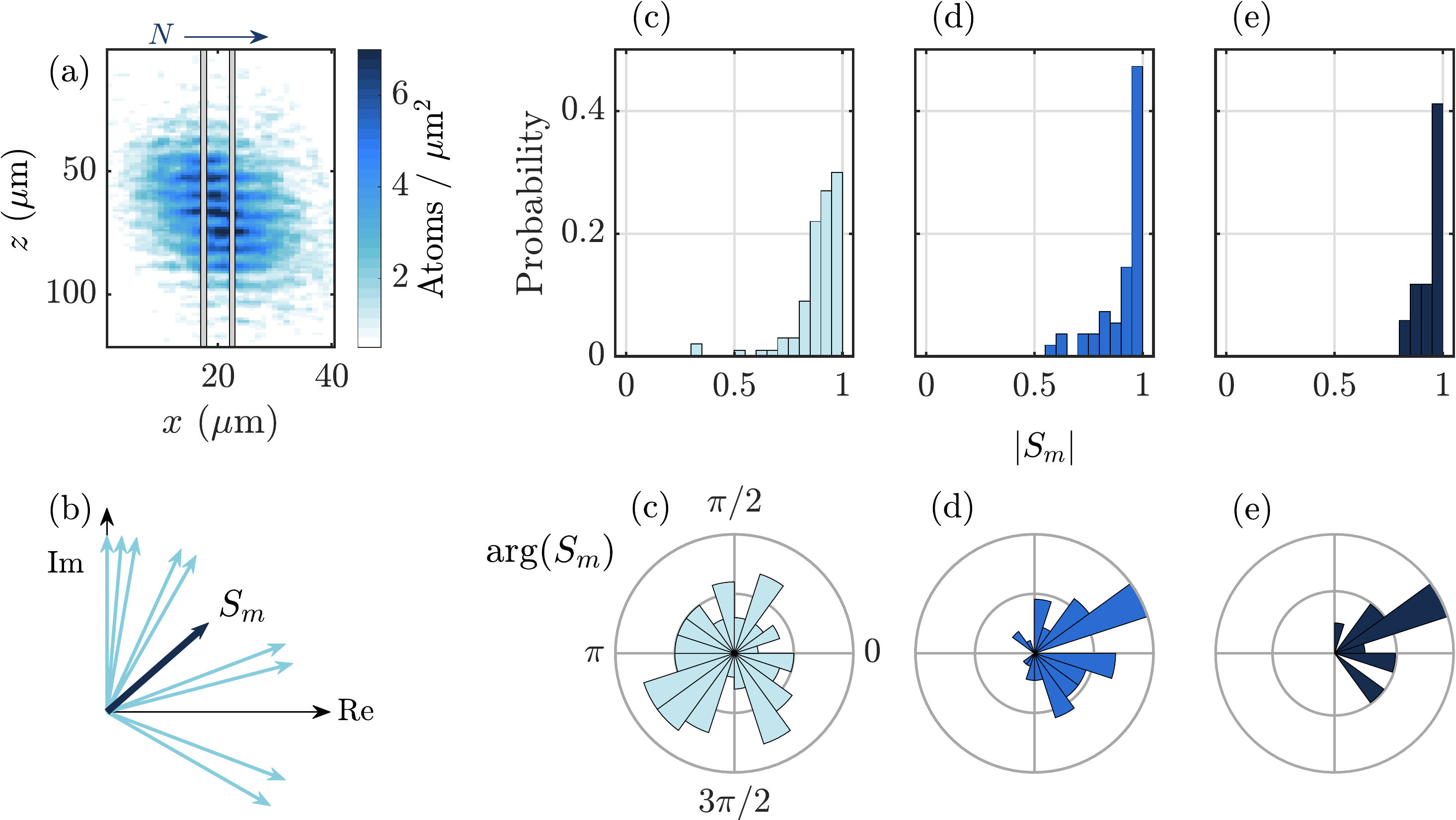}
	\caption[]{Analysing the coherence of the splitting process for \gls{3D} Bose gases. (a)~We extract the local relative phase $\Phi_x$ from pixel column $x$ using equation~\ref{eq:fit}. Two values of $x$ are indicated by the grey boxes. The absorption image shown is typical of the interference patterns for large barrier heights. (b)~Cartoon of the phasor analysis of the local phases $\{ \Phi_x \}_m$ (light blue) to produce a resultant phasor $S_m$ (dark blue) from which we extract the magnitude and argument. (c-e, upper) Distribution of $|S_m|$ across at least 30 experimental sequences for barrier heights $\Delta E$ of \SI{3.6}{\kilo\hertz}, \SI{2.2}{\kilo\hertz} and \SI{1.5}{\kilo\hertz}. (c-e, lower) Distribution of global phase differences arg($S_m$) for the corresponding data. We hold the \glsplural{BEC} for \SI{50}{\milli\second} before imaging the distribution after \SI{21}{\milli\second} of \gls{TOF}. Different axes limits for (c-e, lower) are chosen to highlight the distributions.}
	\label{fig:Phases}
\end{figure}

There are two measures of coherence that can be extracted from our fringe patterns, and our analysis is illustrated in figure~\ref{fig:Phases}.
The first of these refers to the local relative phase $\Phi_x$ across a single sample, where coherence is characterised by straight interference fringes which occur when $\Phi_x$ is constant.
To analyse the straightness, we fit equation~\ref{eq:fit} to the fringe pattern at $N=9$ positions along the $x$-axis, as illustrated in figure~\ref{fig:Phases}~(a), to extract a set of local relative phases $ \{ \Phi_x \}_m$ for image $m$.
Using the method of phasors, we evaluate the resultant $S_m = \sum_x \mathrm{exp}(i \Phi_x ) / N$, for the set $\{ \Phi_x \}_m$, as shown in figure~\ref{fig:Phases}~(b).
For straight fringes, the phasors align and $|S_m| \sim 1$, whereas wavy or random fringes produced by strongly-varying local relative phases lead to $|S_m| \to 0$.
Coherence in the local relative phase is necessary for the investigations proposed in reference~\cite{Mathey2010,Mathey2017}.
The fringes must be straight immediately after the splitting procedure, to allow differences between the decoupled clouds arising from subsequent dynamics or the presence of phase fluctuations to be distinguished.

The second measure of coherence that may be extracted from the fringe patterns is the variation in the mean local phase arg($S_m$) (i.e. the global phase difference) from one experimental sequence to the next~\cite{Schumm2005}.
This quantity is only meaningful for flat fringes.
Coherence of the global phase difference between successive sequences is important for precision measurements, such as gravimetry, where external fields are measured by their effects on the global phase difference and are consistent across the whole sample.

We repeat the series of measurements for a range of barrier heights $\Delta E$, where the peak of the barrier is $\Delta E/h =$ \SI{3.6}{\kilo\hertz}, \SI{2.2}{\kilo\hertz}, \SI{1.5}{\kilo\hertz} above the potential energy of the wells, as determined from a numerical simulation of the potentials.
Figure~\ref{fig:Phases}~(c-e, upper) illustrate the distributions of $|S_m|$ across at least 30 separate experimental sequences as a function of $\Delta E$.
For all barrier heights, we observe values of $|S_m|$ close to 1, indicating approximately straight fringes for all experimental parameters tested, which is to be expected for \glsplural{BEC} in the \gls{3D} regime which have a constant phase across the condensate.
Figure~\ref{fig:Phases}~(c-e, lower) displays the corresponding distributions of arg($S_m$).

We observe that fluctuations in the mean relative phase are smaller for lower barrier heights.
This could arise via a number of mechanisms, such as: phase-locking via tunnel coupling; phase-locking via collisions with thermal atoms that are sufficiently energetic to traverse the barrier; or amplitude fluctuations in the \rf{} fields.
Additional work to determine the mechanism responsible is ongoing~\cite{Barker2020b}.

\section{Interference of two-dimensional Bose gases}
\label{sec:2D}

\begin{figure}
	\centering
	\includegraphics[width = 0.9\linewidth]{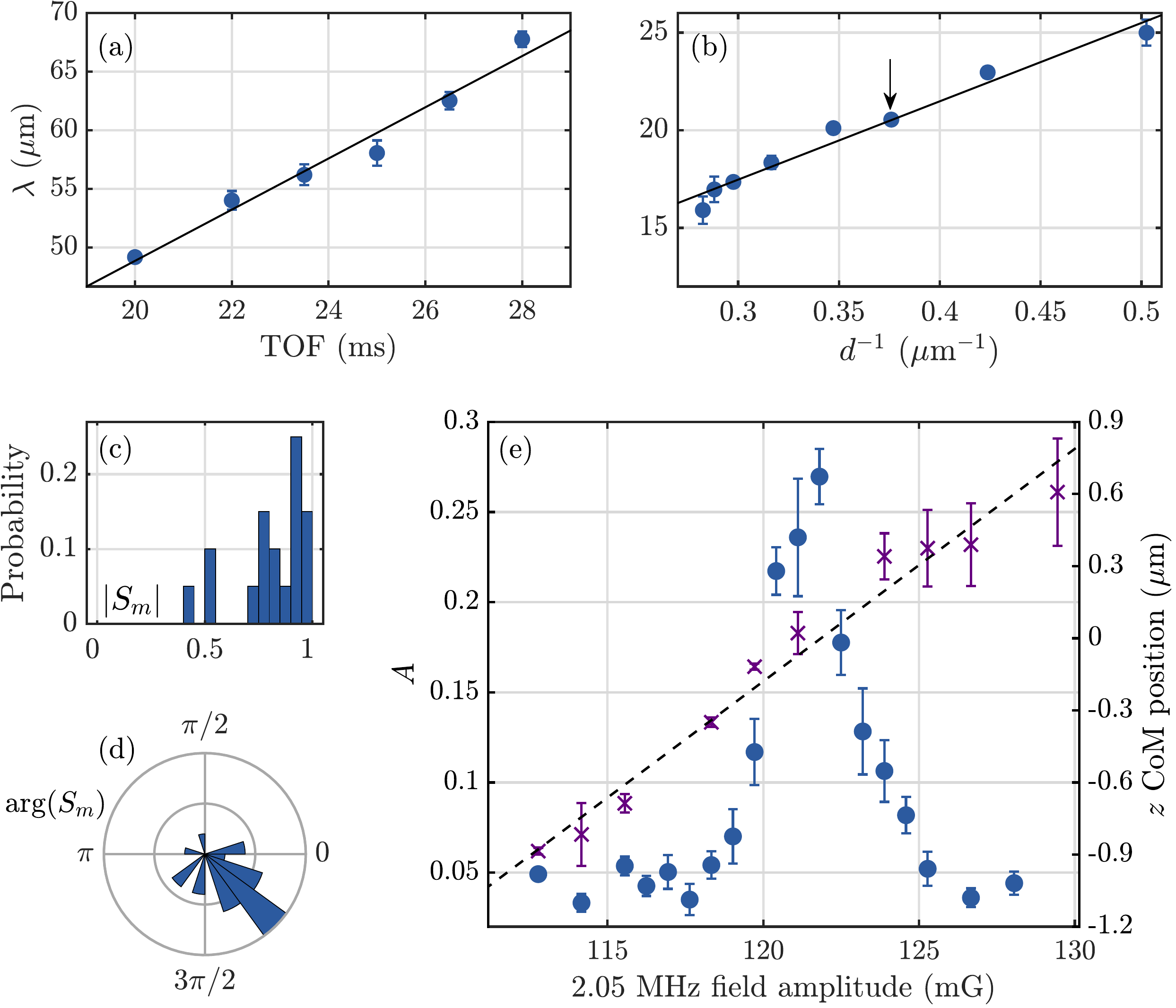}
	\caption[Tuning the contrast of MWI via the relative well populations.]{\gls{MWI} with \gls{2D} Bose gases. (a) \& (b) Fringe spacing $\lambda$ as a function of $t_{\mathrm{TOF}}$ and $d^{-1}$, respectively. The well separations in (b) are determined from a numerical simulation of the potential. (c)~\&(d) $|S_m|$ and arg($S_m$) for the 20 measurements marked by the arrow in (b). (e) Variation of the contrast $A$ of \gls{MWI} (blue) as a function of the amplitude of the \rf{} field for the lower well. The centre-of-mass position along the $z$-axis for a measurement of the \textit{in situ} distribution is also shown (purple), with a linear fit as a guide to the eye (dashed). }
	\label{fig:ContrastCoM}
\end{figure}

The trap frequency in the vertical direction for each well can be increased by reducing the amplitude of the dressing \rf{} fields and increasing the quadrupole field gradient~\cite{Merloti2013}.
We apply \rf{} fields at \SIlist{1.95;2;2.05}{\mega\hertz}, with a magnetic field gradient of \SI{165}{\gauss\per\centi\metre}; the closely spaced \rf{} frequencies produce a well separation of approximately \SI{2}{\micro\metre}.
We load roughly \SI{1.5d4} atoms into each well of the double-well potential and experimentally verify the frequencies of axial oscillation $\omega_z / 2 \pi$ to be \SI{1.12}{\kilo\hertz} (upper) and \SI{0.95}{\kilo\hertz} (lower).
The radial trap frequencies in both wells are measured to be \SI{27}{\hertz}.
The well separation is verified by variation of the $t_{\mathrm{TOF}}$ to be 2.1(0.1)~\si{\micro\metre}, as shown in figure~\ref{fig:ContrastCoM}~(a); we also vary the well separation via the barrier \rf{} amplitude (figure~\ref{fig:ContrastCoM}~(b)).
In this investigation, however, we use barrier heights of $\Delta E / h \ge $\SI{2}{\kilo\hertz}, to avoid a trivial phase-locking scenario in which the energy of the ground state of trapped gases exceeds the barrier height.

We produce trapped gases in the quasi-\gls{2D} regime, where the axial dynamics are frozen out and atoms occupy the lowest harmonic oscillator eigenstate in the axial trapping potential~\cite{Hadzibabic2009,Merloti2013,DeRossi2017}.
This occurs when the thermal energy $k_B T$ and chemical potential $\mu$ of the gases are much smaller than the axial harmonic oscillator spacing $\hbar \omega_z$.
In our case, $\mu/\hbar \omega_z=0.51(0.04)$ and $k_B T/\hbar \omega_z=0.54(0.09)$, for a temperature $T = 35(4)~\si{\nano\kelvin}$; the temperature is extracted from the thermal wings of the density distribution along the $x$-axis at several durations of \gls{TOF} expansion.

Figure~\ref{fig:ContrastCoM}~(c) shows that the splitting procedure produces straight fringes, with $|S_m| \sim 1$ across several experimental cycles, while (d) demonstrates that the mean relative phase is repeatable; this data corresponds to the series of measurements marked with the arrow in (b).
To our knowledge, this work constitutes the first realisation of coherent splitting of \gls{2D} Bose gases in a magnetic trapping potential.
Furthermore, the flat fringes show that our splitting procedure does not produce local variations in the relative phase, which is a critical step towards investigations of the out-of-equilibrium dynamics of \gls{2D} systems~\cite{Mathey2010,Mathey2017,Luksch}.
Next, we discuss methods to balance the double well and to maximise the \gls{MWI} contrast.

The relative population of atoms in each well affects the contrast of the \gls{MWI} pattern: maximal contrast is achieved by having an approximately balanced distribution of atoms across the two wells.
This can be achieved in our apparatus via control of the relative amplitudes of the \rf{} dressing fields which form the upper and lower wells~\cite{Harte2018}.
For example, decreasing the amplitude of the \rf{} component which forms a given well lowers the well's potential energy, and increases the number of atoms loaded into that well during the splitting procedure.
Figure~\ref{fig:ContrastCoM}~(e) illustrates the contrast $A$ as we vary the amplitude of the \SI{2.05}{\mega\hertz} \rf{} field.
We also illustrate the centre-of-mass position along the $z$-axis for the \textit{in situ} spatial distribution of atoms prior to \gls{TOF} expansion.
The variation in contrast clearly shows a peak where the population is expected to be balanced, which is corroborated by the centre-of-mass position being roughly half-way between the two wells (the lower well is centred on approximately \SI{-1}{\micro\metre} and the upper well at \SI{1}{\micro\metre}).
For this choice of experimental parameters, around the central optimum, we find the contrast reduces by approximately 0.07 per \si{\milli\gauss} drift in \rf{} field amplitude, which is a typical level of \rf{} amplitude control.

While, for the double-well potential shown here, the change in centre-of-mass is large enough to resolve optically, this strategy will fail for smaller separations that are below the diffraction limit of the imaging system.
In that case, the wells may still be balanced by maximising the fringe contrast, as we show here.

The maximum contrast in our measurements shown so far is limited to $0.3$, which we attribute to the effects of integration over the thermal component within the gas. 
This can be avoided using a method of selective imaging, as we now discuss.

\section{Selective imaging}
\label{sec:Tomo}

In our apparatus, absorption imaging is performed by first `repumping' atoms from the \Fo{} hyperfine level, where $F$ is the total angular momentum quantum number, to the \Ft{} level.
We then image the atomic distributions using light resonant with a transition from the \Ft{} level to an excited state.
When the entire cloud is repumped, all atoms along a column parallel to the imaging axis $\hat{\mathbf{e}}_y$ absorb light and contribute to the optical density.
This also includes the thermal component, which does not contribute to the interference and so reduces the contrast of the fringe pattern.
Furthermore, fluctuations in the phase of the fringes perpendicular to the imaging plane also act to reduce the contrast.

\begin{figure}
	\centering
	\includegraphics[width = 0.96\linewidth]{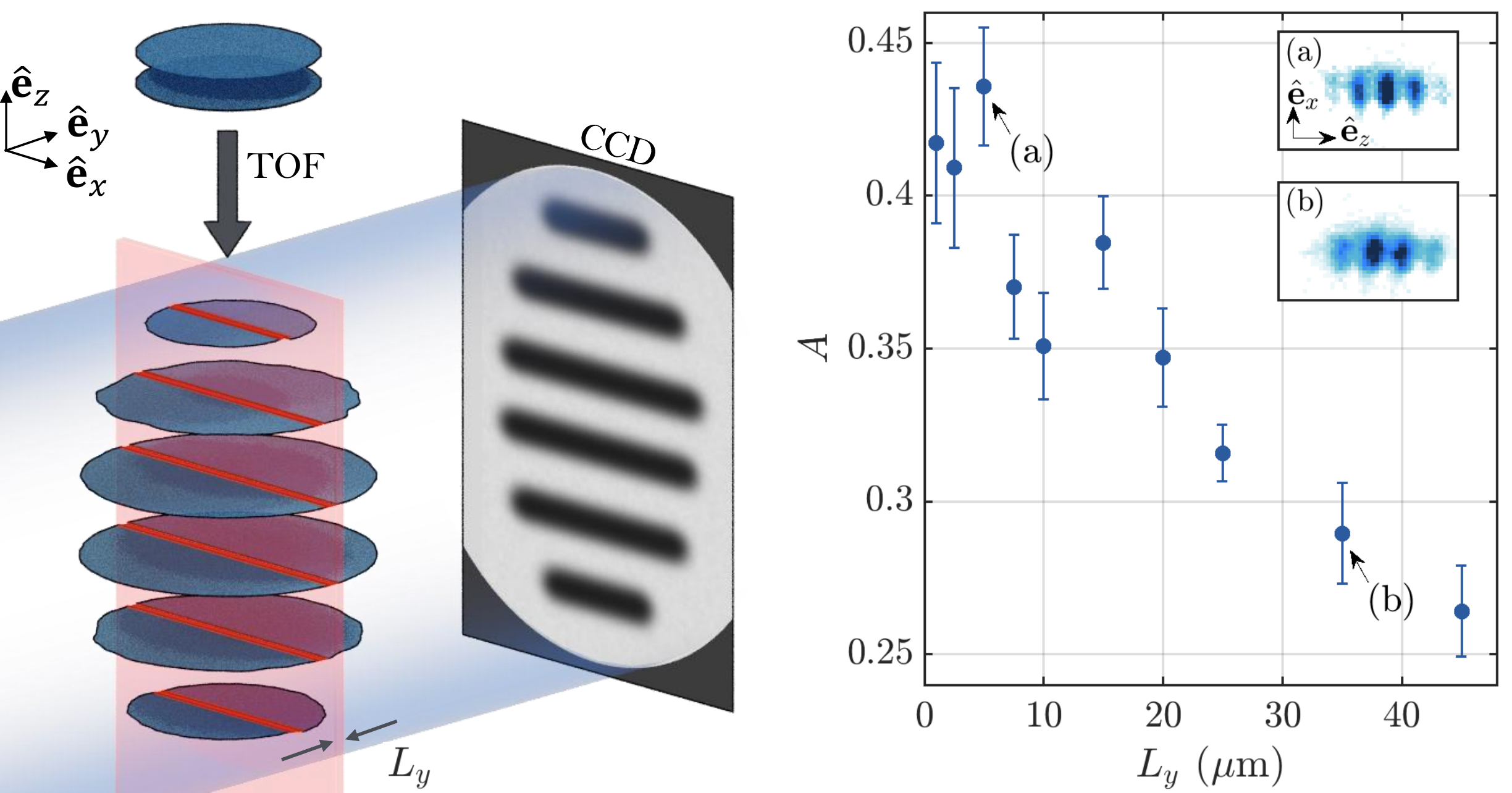}
	\caption[Tomographic imaging and MWI.]{Left: schematic of the selective repumping sequence. We begin with \glsplural{BEC} trapped in a double-well potential before performing \gls{TOF} expansion. Atoms in a state with $F=1$ are optically pumped to $F=2$ by a sheet of repumping light with thickness $L_y$ (red). We then image atoms with $F=2$ using resonant light (blue), producing an absorption image captured by a CCD sensor. This figure is adapted from reference~\cite{Luksch}. Right: fringe contrast $A$ vs. the thickness of the repumping sheet $L_y$. The width of the atomic distribution is approximately \SI{40}{\micro\metre}. Inset (a)-(b): absorption images taken with $L_y = \SI{5}{\micro\metre}$ and \SI{35}{\micro\metre}, respectively.}
	\label{fig:Repumping}
\end{figure}

Instead of imaging the entire cloud, we selectively image a thin sheet of the atomic distribution, here implemented using a \gls{DMD}.
This device comprises an array of mirrors which can be independently controlled to realise arbitrary intensity patterns and is widely used in the cold-atom community to form optical trapping potentials~\cite{Gauthier2016,Muldoon2012}.
The \gls{DMD} is used in combination with a vertical lens to selectively repump only atoms in a thin sheet of controllable thickness $L_y$ (red), normal to the imaging direction, before imaging the atoms using resonant light (blue).
Thus the absorption image captures only the integrated density over distance $L_y$, rather than the full width of the cloud.
Our approach is illustrated in figure~\ref{fig:Repumping}~(left). 
The high visibility interference fringes in reference~\cite{Andrews1997} were obtained by a similar technique, however, the optical apparatus in that case was fixed and did not offer the tunability of a \gls{DMD}.

Using the same analysis methods as presented in section~\ref{sec:Coherence}, the full distribution of the contrast $A(L_x, L_y, t)$ can be constructed, where $t$ indicates the hold time before \gls{TOF} and $L_x$ is the integration length along $\hat{\mathbf{e}}_x$.
These methods permit the future investigation of phase correlations within the gas following a quench~\cite{Mathey2010}.

\section{Outlook}
\label{sec:Conclusion}

The experimental methods described in this paper provide powerful tools for the investigation of the phase distributions of quantum gases, in particular the \gls{BKT} transition in \gls{2D} systems. 
We have shown that our methods produce straight fringes, indicating that the relative phase between the daughter clouds is spatially homogeneous following the splitting procedure.
This is a necessary initial condition to study the subsequent time evolution of correlations within the gas, which can be extracted from analysis of the contrast of the \gls{MWI} pattern~\cite{Polkovnikov2006}.
The interference contrast can be related to the first-order correlation function $g_1(x) = \langle \hat{\Psi}^{\dagger}(x) \hat{\Psi}(0) \rangle$ of the sample, where $\hat{\Psi}(x)$ is the bosonic annihilation operator at position $x$, to discriminate between the superfluid and normal phases of a \gls{2D} gas~\cite{Polkovnikov2006,Hadzibabic2009}.
In the normal phase, $g_1(x)$ is an exponentially decaying function and the squared-contrast, integrated over $L_x$, is predicted to decay with $A^2 (L_x) \propto L_x^{-1}$. 
In the superfluid state, $g_1(x)$ decays slowly with an algebraic functional form, such that $g_1(x) \propto x^{-\alpha}$, and leads to $ A^2 (L_x) \propto L_x^{-2\alpha}$.
The evolution of the gas from superfluid to normal phase following the quench can therefore be probed via the functional form of the integrated squared-contrast, to elucidate the timescales of relaxation and equilibration~\cite{Mathey2010}.
Selective imaging using the \gls{DMD} avoids extensive integration of the fringe pattern along $L_y$ which would otherwise reduce the peak contrast, both via the contribution of the coexisting thermal component as well as integration over phase fluctuations which may be present in the clouds.

For the investigation of out-of-equilibrium phenomena it is desirable to have isolated daughter clouds that evolve completely independently. 
The multiple-\rf{}-dressed potentials may also be engineered to retain some coupling between the daughter clouds.
This provides a means to investigate the time-dependent dynamics of the coupled system or to realise the \gls{2D} sine-Gordon Hamiltonian, both in and out of equilibrium, which is a subject of widespread theoretical interest~\cite{Bratsos2007,Benfatto2007,vanNieuwkerk2018}.
The closest well separation of approximately \SI{2}{\micro\metre} that we have obtained is similar to experiments elsewhere that observe tunneling dynamics~\cite{Luick2019,Pigneur2017,Albiez2005}.

\section*{Acknowledgements}
We thank Thomas Bailey for assistance with the implementation of the DMD.
This work was supported by the EPSRC Grant Reference EP/S013105/1. 
We gratefully acknowledge the support of NVIDIA Corporation with the donation of the Titan Xp GPU used for parts of this research. 
AJB, DG, AB and KL thank the EPSRC for doctoral training funding.

\section*{References}

\bibliographystyle{iopart-num}
\bibliography{library}

\end{document}